\documentclass[twoside]{dis08}
\usepackage[latin1]{inputenc}
\usepackage[dvips]{graphicx,epsfig,color}
\usepackage{wrapfig,rotating}
\usepackage{amssymb,amsmath,array}

\begin{document}

\title{Implications of CTEQ PDF analysis\\ 
for collider observables\thanks{
Contribution to the Proceedings of XVI International Workshop on
Deep-Inelastic Scattering and Related Subjects (DIS 2008), London,UK,
7-11 April, 2008}
}

\author{Pavel M. Nadolsky$^{1,2}$\\
\vspace{.3cm}\\
%
1- Department of Physics, Southern Methodist University,
Dallas, TX 75275, U.S.A.\\ \vspace{.1cm}
2- Department of Physics and Astronomy, Michigan State University,
East Lansing, MI 48824, U.S.A.
}

\maketitle

\begin{abstract}
In the context of a recent CTEQ6.6 global analysis, we review 
a new technique for studying correlated theoretical uncertainties in 
hadronic observables associated with imperfect knowledge of parton
distribution functions (PDFs). The technique is based on the computation
of correlations between the predicted values of physical observables 
in the Hessian matrix method. It can be used, for
example, to link the dominant PDF uncertainty
in a hadronic cross section to PDFs for individual parton flavors
at well-defined $(x,\mu)$ values. As an illustration, we apply the PDF
correlation analysis to study regularities in the PDF dependence of 
$Z$, $W$, and Higgs boson
production cross sections at the Tevatron and LHC.
\end{abstract}

\vspace{\baselineskip}
Theoretical description of LHC observables requires 
accurate parton distribution 
functions (PDFs), determined from a
comprehensive fit of theoretical cross sections to a diverse range 
of experimental data. Reliability of the existing PDF parametrizations
depends on our understanding of
 rich connections existing between PDFs of different flavors
and in different kinematical ranges, arising as a
consequence of physical symmetries (such as scale invariance or parton
sum rules) and experimental constraints implemented in the global fit. 
This talk addresses the
need to explore such connections effectively 
with the help of new quantitative tools provided by global PDF
analyses. I will refer to the slides of the talk available at \cite{NadolskyDIS1}.

A convenient measure of the relation between the PDF dependence of
two physical quantities $X$ and $Y$ is the angle $\varphi$ formed by the
gradient vectors $\nabla X$ and $\nabla Y$ in the space of $N$ PDF
parameters ${a_i}$. The correlation angle was
originally introduced in \cite{Pumplin:2001ct,Nadolsky:2001yg} and
systematically explored in the context of the recent CTEQ6.6 NLO PDF analysis 
\cite{Nadolsky:2008zw}. 
Together with the usual PDF uncertainties $\Delta X$ and $\Delta Y$, 
$\varphi$ can be derived using the
Hessian matrix method \cite{Pumplin:2001ct} from the values ${X_i^{\pm},
  Y_i^{\pm}}$ ($i=1, N$) taken by $X$ and $Y$ for maximal
positive ($+$) and negative ($-$) displacements for each 
PDF eigenparameter $a_i$ within the fit's tolerance region. 
The cosine of $\varphi$ is explicitly given by
\begin{displaymath}
\cos\varphi=\frac{1}{4\Delta X\,\Delta
  Y}\sum_{i=1}^{N}
\left(X_{i}^{(+)}-X_{i}^{(-)}\right)\left(Y_{i}^{(+)}-Y_{i}^{(-)}\right).
\end{displaymath}

The usefulness of $\cos\varphi$ can be appreciated by noticing
that $\Delta X$, $\Delta Y$, and $\cos\varphi$ 
are sufficient to establish a Gaussian probability distribution
$P(X,Y|\mbox{data})$ for finding certain values of $X$ and $Y$
based on the experimental data sets included in the global
analysis. Hence, the three parameters come in handy in certain
statistical estimates. For example, they determine joint confidence 
regions for the $X$-$Y$ pair 
(error ellipses in the $X$-$Y$ coordinate plane), such as the error
ellipses for $W$ and $Z$ production cross sections discussed below.
They are also sufficient for the estimation of the PDF uncertainty 
$\Delta f$ of any function $f(X,Y)$ according to the formulas 
presented in \cite{Nadolsky:2008zw}. 
 
The value of $\cos\varphi$ is a quantitative measure of 
our ability to reduce the
PDF uncertainty in $Y$ by precisely measuring $X$. The measurement of
$X$ would constrain $Y$ substantially if $X$ and $Y$ 
are strongly correlated ($\cos\varphi \approx 1$) 
or anticorrelated ($\cos\varphi
\approx -1$). Conversely, if $\cos\varphi
\approx 0$, the measurement of $X$ is not likely to constrain $Y$.

Some applications of the correlation
analysis  were considered in \cite{Nadolsky:2008zw}. We focus, in
particular, on correlations between PDFs of specific flavors and
physical cross sections. The recent CTEQ6.6 PDFs with more precise
treatment of $s$, $c$, and $b$ quarks (summarized
in \cite{NadolskyDIS2}) allow us to assess the flavor
dependence of the PDF-induced correlations as reliably as possible. 

An instructive example of PDF-induced correlations is provided 
by total cross sections for $Z$, $W$, and $t\bar t$ production 
($\sigma_Z$, $\sigma_W$,
and $\sigma_{t\bar t}$) at the LHC. These cross sections are
plotted pairwise as dots for 41 CTEQ6.1 PDF sets in two figures on
slide 3. In the upper figure, the dots for 41 pairs of $Z$ and $W$ cross 
sections lie within a narrow ellipse, with the center corresponding to
the best-fit CTEQ6.1M PDF set. For each extreme PDF set, 
variations in $\sigma_{Z}$ and $\sigma_{W}$ tend to be of 
the same sign and of similar relative magnitudes, 
indicating a strong correlation in their PDF dependence. 

On the other hand, variations due to the PDFs in the $t\bar t$ total cross
section (lower figure) tend to be opposite 
in sign to those of $W/Z$ cross sections, indicating a substantial
anticorrelation \cite{Campbell:2006wx}. The two
figures or simple arguments do not explain 
what drives the (anti-)correlation, for instance,
why the $W$ and $Z$ cross sections
(dominated by light-quark scattering) are
anticorrelated with the $t\bar t$ cross section (dominated by $g-g$
scattering). 

In order to reveal the underlying physics mechanism, on slides 14a,b 
we plot $\cos\varphi$ between the $Z$ boson production cross
sections and
PDFs $f_{a}(x,Q)$ of different flavors, evaluated for the Tevatron
Run-2 and LHC as a function of
the momentum fraction $x$ at an energy scale $Q=85$ GeV. The results
for $W$ boson production are qualitatively the same \cite{Nadolsky:2008zw}.
A PDF flavor 
having a strong correlation with $\sigma_Z$ contributes a large part of
the PDF uncertainty $\Delta\sigma_Z$ in $\sigma_Z$. Additional constraints
on this flavor would help reduce $\Delta \sigma_Z$. 
In $Z$ boson production, 
the largest correlations occur at momentum fractions $x$
of order $M_{Z}/\sqrt{s}$, i.e., at $x\sim$0.05 at the Tevatron and 0.007 at the LHC,
 corresponding to central rapidity production.

According to the figures, correlations 
in $\sigma_Z$ at the LHC are not the same 
as at the Tevatron. At the Tevatron (slide 14a), 
large correlations ($\cos\varphi
\approx 0.95$) exist with $u$, $\bar{u},$ $d,$ and $\bar{d}$
PDFs, while no tangible correlation occurs
with the PDFs of other flavors. At the LHC (slide 14b),
the largest correlations are driven by charm, bottom, and gluon PDFs,
followed by smaller correlations with $u,$ $d,$ and $s$ quarks.
This feature may come across as surprising, as  $Z$ bosons 
are mostly produced in $u$ and $d$ quark-antiquark scattering 
at both colliders. However, this dominant channel contributes little 
to the PDF {\it uncertainty} at the LHC because of tight constraints 
imposed on the $u$ and $d$ PDFs at relevant $x$  
by the DIS and Drell-Yan data. Rather, the bulk of the PDF
uncertainty comes from the less constrained $s,
c, b,$ and $g$ scattering channels.

In the LHC case, a large positive correlation of $W$, $Z$ cross sections
with $g,$ $c$, and $b$ PDFs at $x\sim0.005$
is accompanied by a large anticorrelation ($\cos\varphi\sim-0.8$)
with the same PDFs at $x\sim0.1-0.2$. The anticorrelation 
reflects the nucleon's
momentum sum rule, which demands that variations in the gluon PDF
at small $x$ are compensated by opposite variations at large $x$ in order 
to satisfy 
\begin{displaymath}
\int_0^1 x f_g(x,Q) dx + \sum_{\mbox{quark flavors}}\int_0^1 x \left[f_q(x,Q)
  + f_{\bar q}(x,Q)\right] dx = 1.
\end{displaymath}
The anticorrelation can be viewed directly at the 
level of the underlying PDFs by
examining $\cos\varphi$ between
$f_{g}(x_1,Q)$ and $f_{g}(x_2,Q)$ as a function of $x_1$ and
$x_2$. A contour plot 
of such dependence 
is shown for $Q=85$ GeV  in slide 13, the upper right figure. 
The colors of the contours correspond to the value of
$\cos\varphi$ according to the palette included in the slide. In this
plot, the red area along the diagonal reflects a trivial perfect correlation
of $f_g (x,Q)$ with itself at the same $x$ ($\cos\varphi=1$ if $x_1 = x_2$).  
The anticorrelation due to the momentum sum rule produces 
dark-blue areas near $(x_1,x_2) = (0.2,0.01)$. 
The same anticorrelation also appears in the case 
of the $c$ and $b$ PDFs (lower figures), the distributions generated 
radiatively from $f_g (x,Q)$  via DGLAP evolution.
It does not occur in the case of light (anti-)quarks. For instance, 
the $u-u$ contour plot in the upper left
figure only shows a weak anticorrelation at $(x_1, x_2)\approx (0.1,
0.7)$ associated with the valence sum rule, $\int_0^1 \left[u(x,Q) -
  \overline u(x,Q)\right] dx =2$. PDF-PDF correlation plots for other
 parton flavors or different $Q$ are posted at \cite{CTEQ66corr}.

Since the gluon anticorrelation originates from a basic sum rule, it universally affects 
processes involving gluon scattering. In particular,
$t\bar t$ production at the LHC is strongly correlated with $f_g
(x,Q)$ at $x\sim 0.1$, and, therefore, anticorrelated with $f_g
(x,Q)$ at $x$ of a few $10^{-3}$ (slide 15). This explains 
why the LHC $t\bar t$ cross sections are anticorrelated with the 
$W,$ $Z$ cross sections. For the same reason, the PDF uncertainty
for Higgs boson production in gluon-gluon fusion at the LHC is correlated
with that for production of $Z$ bosons if the Higgs boson is
relatively light ($M_H = 100-150$ GeV) and strongly anticorrelated if
it is heavy ($M_H \approx 500$ GeV); cf. slide 17.

These findings can be of relevance for various aspects of
the LHC physics program, in view that $\sigma_{Z}$, $\sigma_{W}$, and
$\sigma_{t\bar t}$ will be measured with high precision 
in order to calibrate the LHC experimental
equipment and accurately determine standard-model 
parameters (particularly measure the $W$ boson
and top-quark masses).  Many LHC analyses deal with 
ratios of two cross sections $\sigma_1/\sigma_2$, 
such as those introduced to normalize an LHC
cross section to a well-known ``standard candle''
cross section or deduce statistical significance from the signal
and background event rates. Ratios of correlated (but not
anticorrelated) cross sections have greatly reduced PDF
uncertainty.  The correlation analysis identifies straightforwardly
such pairs of correlated cross sections.  Altogether, the results 
in Ref.~\cite{Nadolsky:2008zw} demonstrate that the
correlation analysis is a simple, yet informative,
technique helping to clarify counterintuitive
aspects of the PDF dependence of collider observables. 

\vspace{\baselineskip}
I thank my coauthors for many discussions of presented results, 
and Jon Pumplin and Wu-Ki Tung for the critical reading of the manuscript.  
This work and participation in the workshop 
were supported in part by the U.S. National
Science Foundation under awards PHY-0354838, PHY-0555545, the U.S.\ Department of Energy
under grant DE-FG02-04ER41299, Lightner-Sams Foundation,
and by the 2008 LHC Theory Initiative Travel Award.

\newpage
\begin{footnotesize}

\end{footnotesize}



\begin{thebibliography}{99}
\expandafter\ifx\csname natexlab\endcsname\relax\def\natexlab#1{#1}\fi
\expandafter\ifx\csname bibnamefont\endcsname\relax
  \def\bibnamefont#1{#1}\fi
\expandafter\ifx\csname bibfnamefont\endcsname\relax
  \def\bibfnamefont#1{#1}\fi
\expandafter\ifx\csname citenamefont\endcsname\relax
  \def\citenamefont#1{#1}\fi
\expandafter\ifx\csname url\endcsname\relax
  \def\url#1{\texttt{#1}}\fi
\expandafter\ifx\csname urlprefix\endcsname\relax\def\urlprefix{URL }\fi
\providecommand{\bibinfo}[2]{#2}
\providecommand{\eprint}[2][]{\url{#2}}

\bibitem{NadolskyDIS1} Slides: 
\verb$http://indico.cern.ch/contributionDisplay.py?contribId=89&confId=24657$

\bibitem{Pumplin:2001ct}
\bibinfo{author}{\bibfnamefont{J.}~\bibnamefont{Pumplin}} \bibnamefont{et~al.},
  \bibinfo{journal}{Phys. Rev.} \textbf{\bibinfo{volume}{D65}},
  \bibinfo{pages}{014013} (\bibinfo{year}{2001}).

\bibitem{Nadolsky:2001yg}
\bibinfo{author}{\bibfnamefont{P.~M.} \bibnamefont{Nadolsky}} \bibnamefont{and}
  \bibinfo{author}{\bibfnamefont{Z.}~\bibnamefont{Sullivan}}
  (\bibinfo{year}{2001}), \eprint{hep-ph/0110378}.

\bibitem{Nadolsky:2008zw}
  P.~M.~Nadolsky,  H.-L. Lai, Q.-H. Cao,
J. Huston, J. Pumplin, D. Stump, W.-K. Tung, C.-P. Yuan,
  Phys.\ Rev.\  D {\bf 78}, 013004 (2008).

\bibitem{NadolskyDIS2} P. M. Nadolsky, arXiv:0809.0945 [hep-ph]; slides:\\
\verb$http://indico.cern.ch/contributionDisplay.py?contribId=248&confId=24657$

\bibitem{Campbell:2006wx}
\bibinfo{author}{\bibfnamefont{J.~M.} \bibnamefont{Campbell}},
  \bibinfo{author}{\bibfnamefont{J.~W.} \bibnamefont{Huston}},
  \bibnamefont{and} \bibinfo{author}{\bibfnamefont{W.~J.}
  \bibnamefont{Stirling}}, \bibinfo{journal}{Rept. Prog. Phys.}
  \textbf{\bibinfo{volume}{70}}, \bibinfo{pages}{89} (\bibinfo{year}{2007}).


\bibitem{CTEQ66corr}
\bibinfo{howpublished}{http://hep.pa.msu.edu/cteq/public/6.6/pdfcorrs/}.

\end{thebibliography}
\end{document}